\documentclass[usenatbib]{mn2e}
\usepackage{natbib}
\usepackage{url}
\usepackage{rotating}
\usepackage{caption}
\usepackage{subcaption}
\usepackage{amsmath}
\usepackage{amsfonts}
\usepackage{amssymb}
\usepackage{graphicx}
\usepackage{float}
\usepackage{aas_macros}
\usepackage{color}

\newcommand\lsim{\mathrel{\rlap{\lower4pt\hbox{\hskip1pt$\sim$}}
        \raise1pt\hbox{$<$}}}
\newcommand\gsim{\mathrel{\rlap{\lower4pt\hbox{\hskip1pt$\sim$}}
        \raise1pt\hbox{$>$}}}

\usepackage{hyperref}
\hypersetup{
    pdfnewwindow=true,      
   colorlinks=true,       
    linkcolor=blue,          
    citecolor=blue,        
    filecolor=blue,      
    urlcolor=blue           
}

\begin{document}

\title[Multi-term periodicity of PG1302]{Multiple periods in the variability of the supermassive black hole binary candidate quasar PG1302-102?}

\author[M. Charisi et al.]{M.~Charisi,$^{1}$\thanks{mc3561@columbia.edu}  I.~Bartos,$^{2}$ Z.~Haiman,$^{1}$ A.~M.~Price-Whelan,$^{1}$ S.~M\'arka$^{2}$\\
$^1$Department of Astronomy, Columbia University, New York, NY 10027, USA \\
$^2$Department of Physics, Columbia University, New York, NY 10027, USA}

\maketitle
\begin{abstract}
\citealt{2015arXiv150101375G} discovered a supermassive black hole
binary (SMBHB) candidate and identified the detected 5.2\,yr period of the
optical variability as the orbital period of the
binary. Hydrodynamical simulations predict multiple periodic
components for the variability of SMBHBs, thus raising the possibility
that the true period of the binary is different from 5.2\,yr. We
analyse the periodogram of PG1302 and find no compelling evidence for
additional peaks.  We derive upper limits on any additional periodic
modulations in the available data, by modeling the light-curve as the
sum of red noise and the known 5.2\,yr periodic component, and
injecting additional sinusoidal signals. We find that, with the
current data, we would be able to detect with high significance (false
alarm probability $<1\%$) secondary periodic terms with periods in
the range predicted by the simulations, if the amplitude of the
variability was at least $\sim$0.07\,mag (compared to 0.14\,mag for
the main peak). A three-year follow-up monitoring campaign with weekly
observations can increase the sensitivity for detecting secondary
peaks $\approx$50\%, and would allow a more robust test of predictions
from hydrodynamical simulations.
\end{abstract}

\section{Introduction}

It is well established that massive galaxies harbour supermassive black
holes in their centres, with masses tightly correlated with global
properties of the host galaxy~\citep{kormendy13}. According to
cosmological models of hierarchical structure formation, galaxies grow
through frequent mergers (\citealt{2005MNRAS.361..776S};
\citealt{2006ApJ...645..986R}). The formation of supermassive black
hole binaries (SMBHB) is a natural consequence of the above
process. Combining the high expected rates of galaxy mergers with the
expectation that SMBHBs spend a large fraction of their lifetimes at
close separation \citep{2002MNRAS.336L..61H}, SMBHBs at sub-parsec
separations should be fairly common, despite the scarcity of
observational evidence.

Recently, \citet[][hereafter G15]{Graham+2015} reported the
detection of strong periodic variability in the optical flux of quasar
PG1302-102. PG1302 is a bright (median V-band magnitude $\sim$15),
radio-loud quasar at redshift $z=0.2784$, with inferred black hole (BH)
mass of $10^{8.3-9.4}M_\odot$. The light curve in optical bands shows
a quasi-sinusoidal modulation, with a best-fit period of
($5.2\pm0.2$)\,yr and amplitude of $\approx$0.14\,mag. G15 suggest
that PG1302 may be a SMBHB at close separation ($\sim$0.01\,pc),
interpreting the observed periodicity as the (redshifted) orbital
period of the binary.

Theoretical work on circumbinary disks predicts that SMBHBs can excite
periodic enhancements of the mass accretion rate that could translate
into periodic luminosity enhancements, not only at the orbital period,
but also on longer and shorter timescales. More specifically, if the
binary is embedded in a thin disk, the gas will be expelled from the
central region due to torques exerted by the binary, creating a cavity
\citep{1994ApJ...421..651A}. Several hydrodynamical simulations
\citep{2007PASJ...59..427H, 2008ApJ...672...83M, 2012ApJ...755...51N,
  2012ApJ...749..118S, 2012A&A...545A.127R, 2013MNRAS.436.2997D,
  Farris+2014} indicate that the interaction of the individual BHs
with the inner edge of the accretion disk can pull gaseous streams
into the cavity, resulting in periodic modulation of the accretion
rate at timescales corresponding to $\approx$1/2 and 1 times the
binary's orbital period. For high BH mass ratios ($q\equiv M_1/M_2
\gsim 0.3$), the cavity is lopsided, leading to the formation of a
hotspot in the accretion disk. The strongest modulation in the
accretion rate is observed at the orbital period of the
overdense region, a few ($\sim$3-8) times the orbital period of the
binary.

These results imply that the observed 5.2\,yr period in PG1302 may not
reflect the orbital period of the binary. If the orbital period of the
binary is shorter/longer than the period of optical valiability 
(hereafter, optical period, $t_{opt}$), there are major
implications for the expected quasar-binary fraction, and the
probability of detecting SMBHBs \citep{2009ApJ...700.1952H}, as well
as the physics of the orbital decay. \citet{Dorazio+2015} showed that,
under the assumption that the optical period corresponds to the longer
period of the hotspot in the accretion disk, PG1302 would be in the
gravitational inspiral regime. This would confirm that SMBHBs can
produce bright electromagnetic emission even at these late stages of
the merger~\citep{2012ApJ...755...51N, Farris+2015}.

The significance of the above considerations motivates us to search
for multiple periodicity in the optical variability of PG1302. In this
paper, we search for secondary peaks in the Lomb-Scargle
periodogram. We assess the statistical significance of the identified
peaks by generating mock time series that show correlated noise, as
expected for quasars \citep{2009ApJ...698..895K}. The false alarm
probability of the most significant secondary peak identified in our
analysis is 6\%, below a reliable detection threshold. We set upper
limits on the amplitude of secondary sinusoid variations that could be
detectable with high significance, in the presence of correlated noise, as well as the already
known $5.2$\,yr periodic signal. Although
the current data do not allow the detection of weak secondary periodic
terms, the addition of three years of observations, can improve the
sensitivity up to a factor of 2, lowering the detection threshold to
$\sim$0.04\,mag.

\section{Data Analysis}
\label{sec:modeling}

We investigate the possibility of multiple periodic terms in the
photometric variability of PG1302 by analysing the light curve
published in G15\footnote{We are grateful to M.~Graham for providing
  us with their calibrated photometric data in electronic form.}. The
light curve consists of data from: (1) the Catalina Real-time
Transient Survey (telescopes CSS and MLS;
\citealt{2009ApJ...696..870D,2011BASI...39..387M}); (2) the Lincoln Near-Earth Asteroid
Research (LINEAR; \citealt{2011AJ....142..190S}) program; (3) the All
Sky Automated Surveys (ASAS; \citealt{1997AcA....47..467P}) and; (4)
other archival data (\citealt{1999MNRAS.309..803G};
\citealt{2000AJ....119..460E}), providing a baseline of
$\sim$20\,yr. The different datasets were calibrated to account for
the differences between the various photometric systems, as detailed
in G15.


For our analysis, we use the generalised version of the Lomb-Scargle
(LS) periodogram
\citep{1976Ap&SS..39..447L,1982ApJ...263..835S,2009A&A...496..577Z}, a
standard method for detecting periodicity in unevenly sampled time
series\footnote{Throughout the analysis, we make use of the astroML
  python package~\citep{astroML,astroMLText}.}. We compute the
periodogram for 1000 trial frequencies on a uniform logarithmic grid,
spanning from $f_{min}=1/T_{data}$ to $f_{max}=1/(2 \Delta T)$, where
$T_{data}$ is the baseline of the light curve and $\Delta T$ is the
median time interval between successive data points.

For each peak in the periodogram, we then calculate the probability
that a peak of equal power can arise by chance (false alarm
probability; hereafter FAP). For this purpose, we generate mock time
series that mimic the optical variability of PG1302. Quasars show
correlated stochastic variability, best described as a damped random
walk \citep{2009ApJ...698..895K}. At high frequencies, the power
spectral density decreases with frequency ($PSD \sim 1/f^2$, or red
noise), whereas, at low frequencies, it becomes flat ($PSD \sim f^0$,
or white noise).  Furthermore, PG1302 exhibits strong sinusoidal
variability with period of $1,884\pm88$\,d (intentified by wavelet and autocorrelation function (ACF) based techniques as detailed in G15) and amplitude of 0.14 mag.

We first identify the best--fit model for PG1302 as follows. We
generate evenly sampled stochastic time series with a dense temporal
resolution ($dt=2\,h$) and peak-to-peak amplitude $A$, and with
a power spectrum $\sim 1/f^2$ \citep{1995A&A...300..707T}. We add a
sine wave, with properties as in G15, and subsample to match the
observation times in the light curve of PG1302. We calculate the
average LS periodogram from 50 realisations. Next, we subtract this
periodogram from the observed LS periodogram of PG1302 and calculate
the residuals.  We repeat the above process, varying the amplitude $A$
between 0.5 to 2 times the peak-to-peak amplitude of the light curve
($m_{max}-m_{min}$) of PG1302, and determine the amplitude $A$ that
minimises the residuals.

With the best--fit amplitude $A$, we generate 10,000 realisations of
red noise with periodic variability, as before, and calculate the LS
periodogram for each realisation. At each trial frequency, we
calculate the FAP by comparing the power in the PG1302's periodogram
to the distribution of power {\em at the same frequency} in the periodograms of the mock
time series. We define FAP$_f$, the false alarm probability {\em per
  frequency} as the fraction of the 10,000 simulated periodograms with
power exceeding the value observed for PG1302.

Based on the results from hydrodynamical simulations discussed above, we
search for secondary peaks
around the main 5.2\,yr peak, i.e. between $(f_{\rm min},f_{\rm
  max})=(2.37, 3.8)$\,nHz and $(9.5, 92)$\,nHz\footnote{These ranges
are illustrated by the shaded areas in Fig.~\ref{Fig:UpperLimits}}. The frequency ranges
together contain a total of $N_{\rm tot}=526$ frequency bins.  Since
these bins have non-trivial correlations, we compute the effective
number of independent bins, $N_{\rm eff}$, for each FAP$_f$, by: (i)
applying the same analysis as above, but replacing the PG1302 data
with each individual simulated mock periodogram, and (ii) determining
the fraction of the 10,000 mock periodograms that show FAP$_f$ (or
lower) at {\em any} frequency between $f_{\rm min}\leq f\leq f_{\rm
  max}$.  For FAP$_f=10^{-4}$, we find $N_{\rm eff}=80$, and the total
FAP$\equiv N_{\rm eff}\times$FAP$_f$=0.008.\footnote{$N_{\rm eff}$ is
  decreased to $\approx 60$ and $\approx 30$ for FAP$_f=10^{-3}$ and
  FAP$_f=10^{-2}$, respectively.}

\section{Results}
\label{sec:results}

If the optical variability of PG1302 is the superposition of multiple
periodic terms, it is expected that significant secondary peaks will
appear in the periodogram. Moreover, following the predictions of
hydrodynamical simulations, we expect the peaks roughly at ratios 1:2
(1 and 1/2 times the orbital period) or from 1:3:6 to 1:8:16 (hotspot
period, orbital period and half of the orbital period of the binary).

More specifically, with the limitations imposed by the currently
available data, we were able to explore three possibilities: (A) If
the optical period $t_{opt}$ is the binary orbital period $t_{\rm bin}$, is there
a secondary peak near $0.5t_{\rm bin}$? (B) If $t_{opt}$ is
$0.5t_{\rm bin}$, is there a periodic component at $t_{\rm bin}$?  (C)
If $t_{opt}$ is the long orbital period $\approx(3-8)t_{\rm
  bin}$ of the hotspot in the disk, as hypothesised by
\citet{Dorazio+2015}, is there a secondary period corresponding to
$\approx 0.5t_{\rm bin}$ or $t_{\rm bin}$? Note that in scenario (A),
periodicity is predicted at $\sim (3-8)t_{\rm bin}$. However, the
20-yr baseline of the light curve does not allow us to probe this
low-frequency range.

\begin{figure}
\includegraphics[height=6cm,width=8cm]{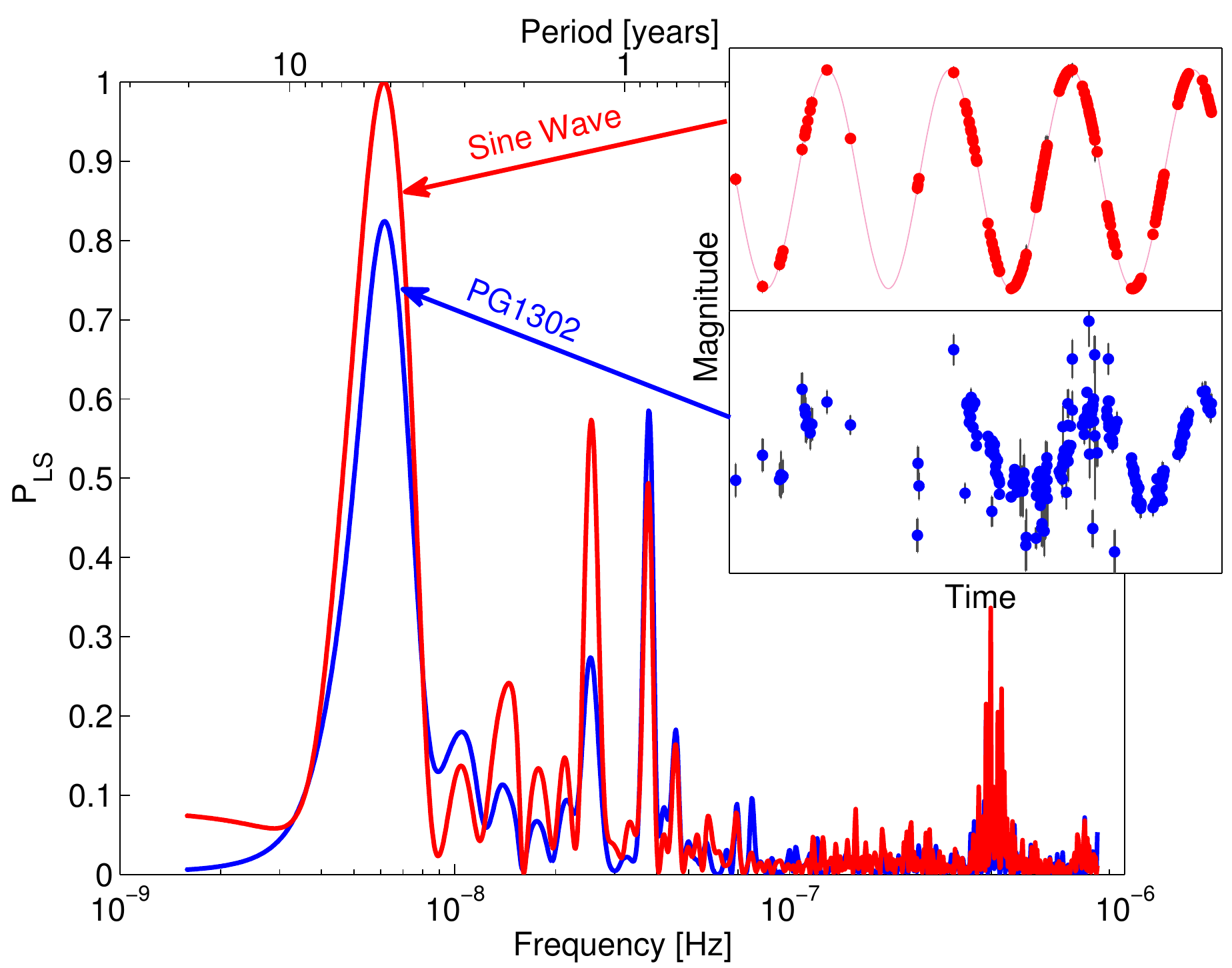}
\caption{LS periodogram of PG1302 (blue) and of a sine wave sampled at
  the observation times of PG1302 (red). The embedded plots illustrate
  the respective light curves, spanning a baseline of
  $\sim$20\,yr. The irregular sampling of the data produces the
  artificial secondary peaks seen in the red curve.}
\label{Fig:SineWave}
\end{figure}

The detection of weaker periodic terms in the presence of a known
significant periodic component is in general challenging. The
periodogram of an unevenly sampled time series is the convolution of
the actual power spectrum of the signal and the window function,
defined by the irregular time sampling. A periodic signal can
introduce aliasing peaks, and hide the secondary peaks.
To illustrate this, in
Fig.~\ref{Fig:SineWave}, we show the LS periodogram of PG1302,
superimposed with the periodogram of a pure sine wave, with period of
5.2\,yr, sampled at the observation times of PG1302. The periodogram
of the sine wave shows a set of artificial secondary aliasing peaks,
which coincide with many of the observed features in the periodogram
of PG1302.\footnote{The aliasing peaks are predictable, and could be
  used to better determine the amplitude and frequency of the
  $5.2$\,yr peak.}

\begin{figure}
\includegraphics[height=6cm,width=8cm]{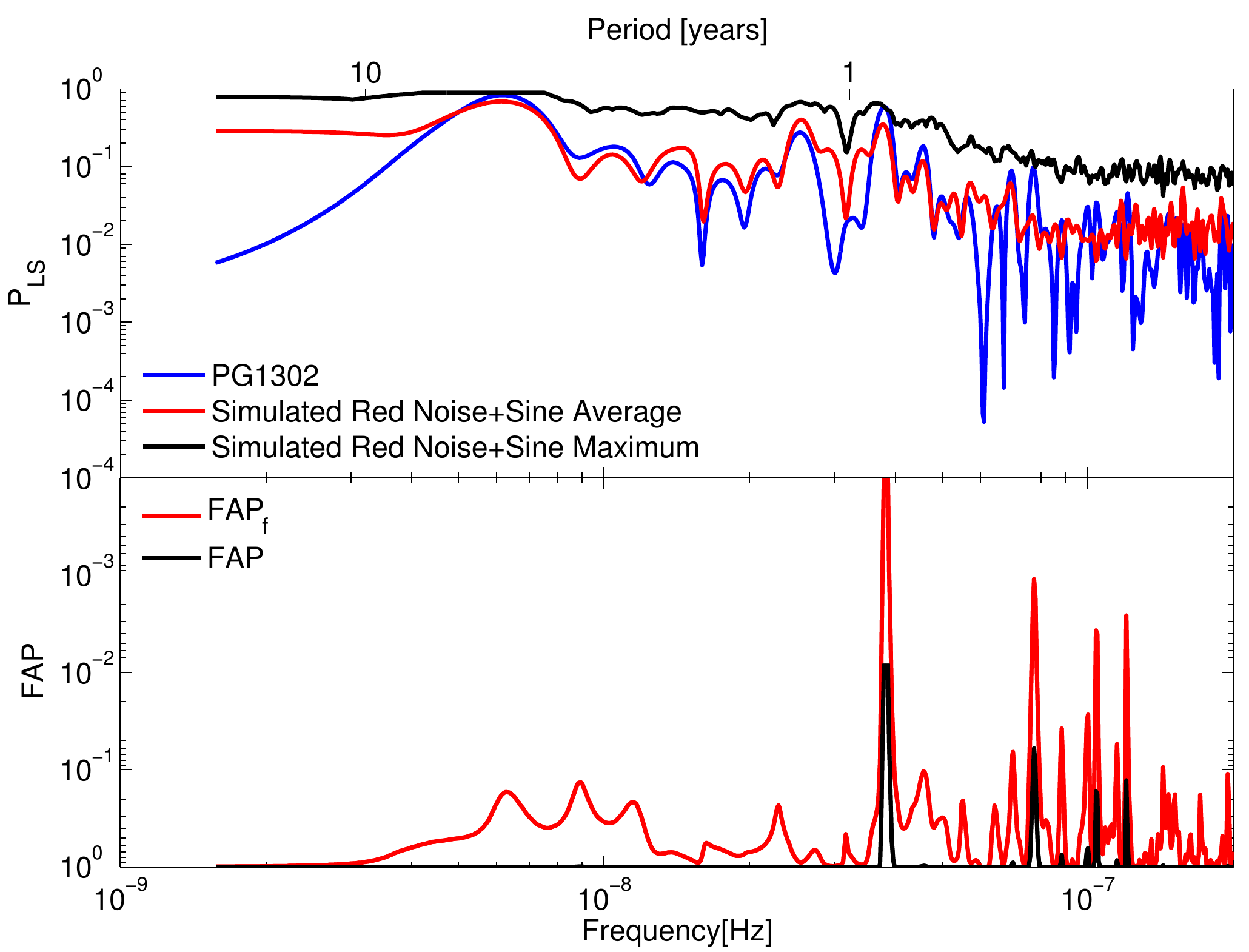}
\caption{{\em Top panel}: LS periodogram of PG1302 (blue),
  superimposed with the average of 10,000 realisations of a model that
  includes stochastic red noise and a 5.2\,yr-sinusoid.  The maximum
  power among these 10,000 realisations is also shown (black),
  corresponding to 1\% false alarm probability threshold. {\em Bottom
    panel}: False alarm probability per frequency (red) and total FAP
  (black) of PG1302's periodogram, as a function of frequency. Note
  that the $y$-axis in the bottom panel decreases upwards.}
\label{Fig:Pvalue}
\end{figure}

We analysed the LS periodogram of PG1302 searching for secondary
peaks, focusing on the frequencies of interest, within a factor of
$\sim$10 around the peak of the main period. We assessed the
statistical significance of the observed secondary peaks using the
method discussed in \S~\ref{sec:modeling} above.
In Fig. \ref{Fig:Pvalue}, we present the periodogram of PG1302 with
the average and the maximum periodogram power from 10,000 realisations
of the simulated variability, with the latter corresponding to our
FAP=1\% significance threshold. In the bottom panel, we show both
FAP$_{f}$ and the total FAP at each frequency. We identify a secondary
peak at $\sim$40\,nHz ($\sim$300\,d), with the lowest FAP ($<0.008$).
This peak is most likely artificial, since it coincides with one of
the strongest aliasing peaks demonstrated in
Fig. \ref{Fig:SineWave}. The most significant peak that does not
overlap with aliasing peaks appears at $\sim$77.5\,nHz
($\sim$155\,d), with a FAP=0.06 (FAP$_{f}$=0.001).

\begin{figure}
\includegraphics[height=6cm,width=8cm]{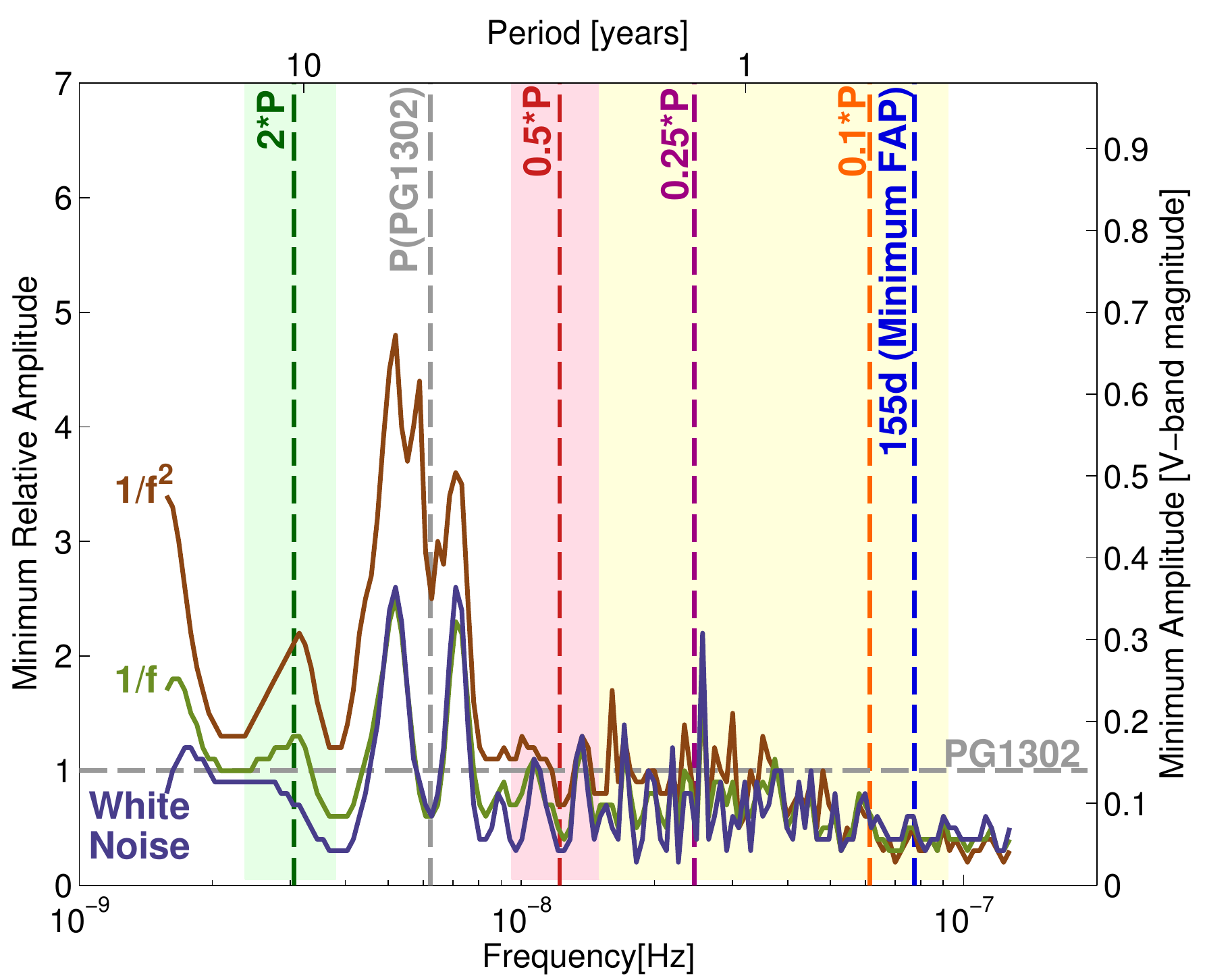}
\caption{The minimum amplitude of a secondary sinusoid term that would
  be detectable at each frequency. The $y$ axis shows the
  amplitude relative to that of the main 5.2\,yr peak (left), and also
  in $V$ mag (right). The three curves correspond to noise models with
  three different power--law power spectra: $\propto 1/f^2$ (brown),
  $\propto 1/f$ (olive), and constant white noise (dark
  blue). Frequency ranges relevant to different scenarios are highlighted with light pink (A), light green (B) and
  light yellow (C). Vertical lines mark specific frequencies of
  interest (see text).}
\label{Fig:UpperLimits}
\end{figure}

We next derive upper limits for putative secondary periodic terms with
the current data. For this purpose, we simulate time series as before,
but with an additional sine wave component. We identified the minimum
amplitude of the second component at which the power in the new peak
in the LS periodogram exceeded the FAP $<1\%$ detection threshold at
least 90\% of the time.  We repeated this process for different
frequencies within the factor of 10 range of interest. In
Fig.~\ref{Fig:UpperLimits}, we present the minimum relative amplitude
(corresponding also to V-band magnitude) of a secondary sinusoid that would
be detectable at different frequencies. The shaded areas in this
figure indicate frequencies relevant to the three possibilities
(A)-(C) discussed above.

As Fig.~\ref{Fig:UpperLimits} shows, we can set weak limits for
scenario~(A); a second periodic component would need a higher
amplitude than the one identified in G15 to be detectable. The reduced
sensitivity in this frequency range is reasonable, since the light
curve is well sampled only in the last $\sim$9\,yr, hindering the
detection of a $\sim$10\,yr period. Moreover, at this frequency range
the effect of the red noise is significant. For scenario (B) and for
the majority of frequencies relevant to scenario (C), the secondary
term needs to have amplitude comparable to the main 5.2\,yr modulation
to be detectable. Finally, for a few specific frequencies relevant to
scenario (C), we can probe secondary sinusoids with amplitudes down to
25\% of the amplitude of the main sinusoid ($\sim$ 0.04\,mag). For the
peak with the lowest FAP (155\,d), the minimum detectable amplitude
would be 50\% of the main sinusoid (0.07\,mag).

\section{Discussion}
\label{sec:discussion}
\subsection{Quasar variability models}

We have thus far modelled the variability of PG1302 using a fixed power
spectrum $1/f^2$. Since quasar variability is described by a damped
random walk (i.e. with the power spectrum flattening at low
frequencies), white (or pink) noise may describe the quasar
variability more accurately at low frequencies. To assess how this
impact our conclusions, we repeated the analysis described above but
with stochastic noise exhibiting $1/f$ (pink noise) and flat (white
noise) power spectra. Our conclusions remain unchanged for these
different noise models. More specifically, we identify the same peak
at $\sim$40\,nH ($\sim$300\,d), which we discard as aliasing with
FAP$<1\%$ both for white and pink noise; we also find the peak at
$\sim$77.5\,nH ($\sim$155\,d) with FAP=0.2 and 0.1 for pink and white
noise, respectively. The upper limits for the different noise models
are also shown in Fig.~\ref{Fig:UpperLimits}. For white noise, the
minimum detectable amplitude is almost constant ($\sim$0.08\,mag) for
all frequencies, but with fluctuations caused by the irregular
sampling similarly to the red noise case.
The results are again similar for pink noise, with the sensitivity falling in-between the red and white noise cases.

\subsection{Historic Data}

Historic photometry of PG1302 is available from the
digitisation of old photographic plates, as part of the Digital Access
to a Sky Century @ Harvard (DASCH; \citealt{2009ASPC..410..101G})
project. The measurements were in unfiltered B photographic magnitude,
which is close to wide-band Johnson B magnitude and were calibrated
with the AAVSO APASS catalog \citep{2012JAVSO..40..430H}.
We extracted the B-band light curve from the online
database\footnote{\url{http://dasch.rc.fas.harvard.edu/lightcurve.php}}
and excluded measurements with high astrometric errors or with
magnitudes within 0.5 of the limiting magnitude, and data points that
were flagged as plate defects. We analysed the periodogram, as before,
assuming red noise variability (without the periodic component) and
finding the best--fit model.

\begin{figure}
\includegraphics[height=6cm,width=8cm]{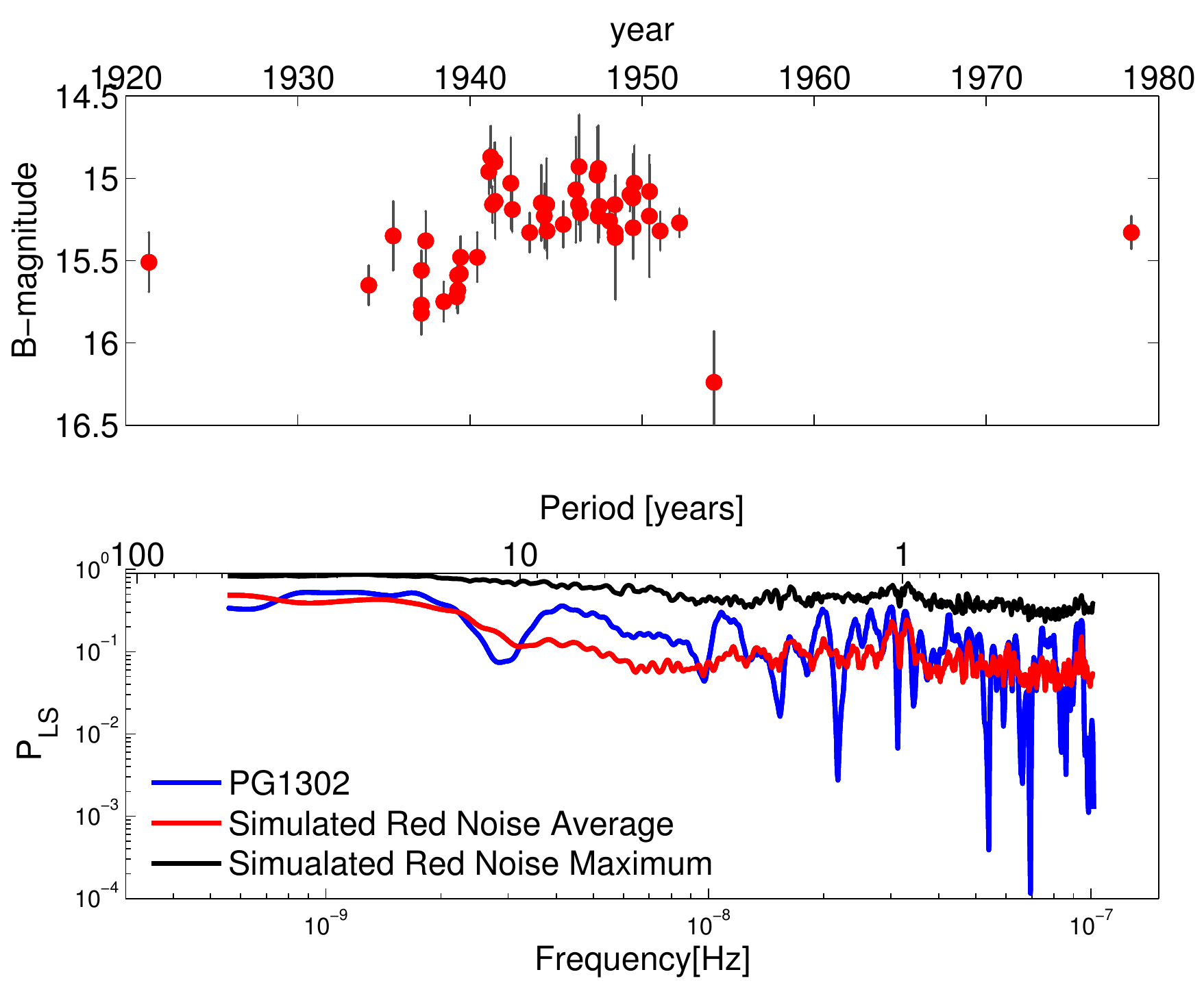}
\caption{{\em Top Panel}: B-band light curve of PG1302 from the DASCH
  project. {\em Bottom Panel}: LS periodogram of the B-band light
  curve (blue), superimposed with the average best fit $1/f^2$ noise
  model (red) and the maximum power of the periodogram from 10,000
  realisations of the noise (black), which corresponds to our
  significance threshold.}
\label{Fig:DASCHData}
\end{figure}

In Fig. \ref{Fig:DASCHData}, we present the B-band light curve and the
LS periodogram, along with the average and maximum power of the LS
periodogram from 10,000 realisations of red noise. All of the
identified peaks (including the 5.2\,yr period from G15) are below the
significance threshold. This result, however, must be viewed in light
of the photometric accuracy of the DASCH dataset. The average
photometric error is 0.18\,mag, exceeding the 0.14\,mag amplitude of
the periodic variability in G15. Interestingly, the two most
significant peaks, within the interval of interest, are observed at
similar frequencies ($\sim$42\,nH and $\sim$76\,nH) as the peaks in
the periodogram from the G15 data, although their significance remains
low.

\subsection{Future Data}

We investigated how the sensitivity for detection of secondary peaks
would increase with three years of weekly follow-up observations. In
order for PG1302 to be observable throughout the year, a network of
telescopes is required (e.g., Las Cumbres
Observatory\footnote{\url{http://lcogt.net/}}). Given that PG1302 is a
bright quasar, even small telescopes can observe it with relatively
small photometric errors. To allow for a flexible, realistic (affected
by weather conditions, resources etc.) observation schedule as well as
to avoid strong aliasing from an exactly periodic time sampling, we
generate a hypothetical follow-up data set, consisting of one randomly
chosen day each week, and allowing an additional scatter of $\pm$3\,h
in the observation times. We also assume photometric errors of
10\,mmag, comparable to the ones in G15.

We analyse the extended light curve as before: we generate 10,000 time
series that exhibit red noise variability with a periodic component as
in G15, sampled at the observation times (including both the existing
and future data) and calculate the LS periodogram. We set our
significance threshold to the maximum power of the periodogram from
the 10,000 realisations (FAP$\leq$1\%). Next, as done above, we
include a secondary sinusoidal term and identify the minimum amplitude
that would result in a peak with power above the threshold 90\% of the
time.

\begin{figure}
\includegraphics[height=6cm,width=8cm]{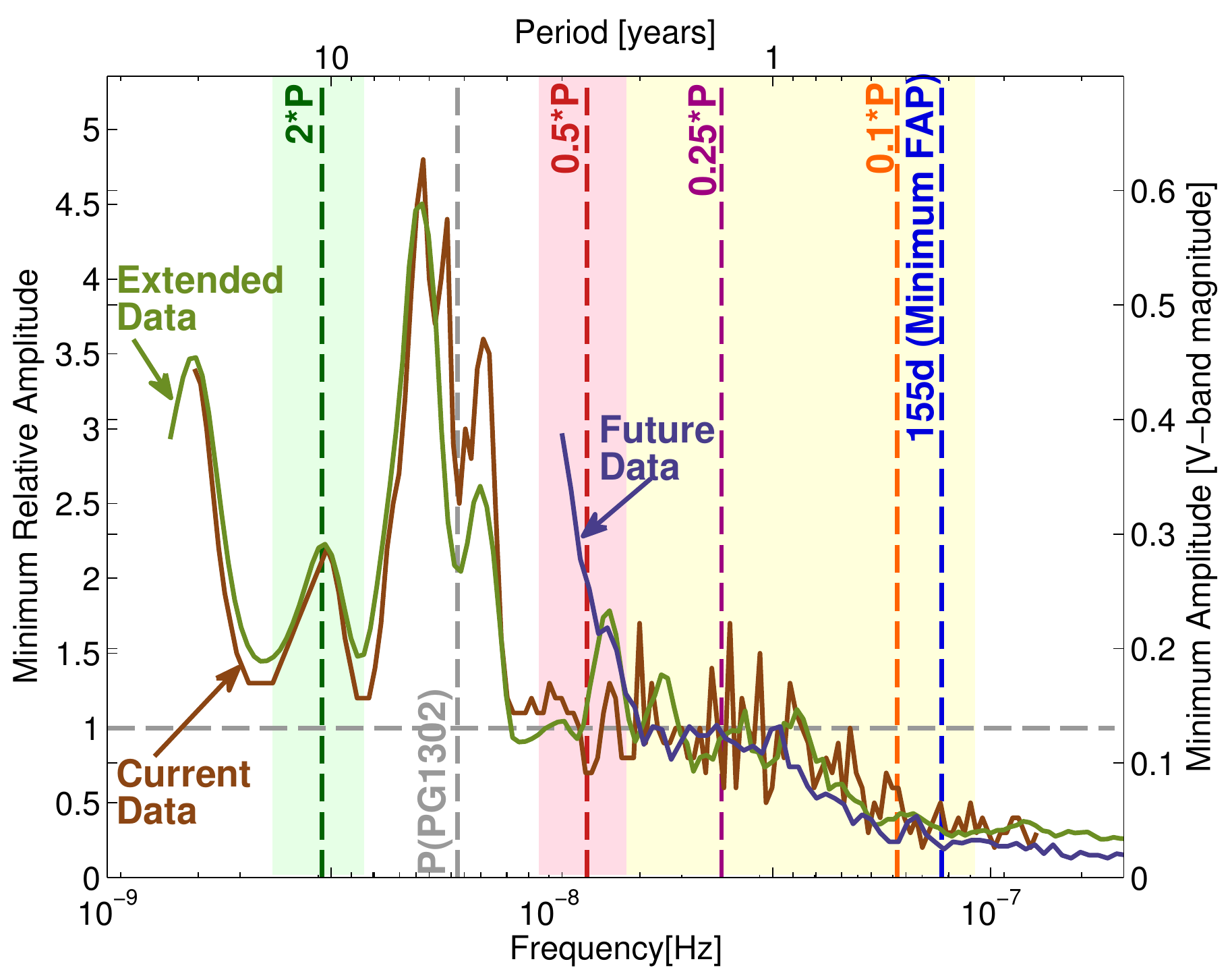}
\caption{Minimum relative amplitude (also in V mag) of a secondary
  sinusoid versus frequency of secondary sinusoid that could be
  detectable in the current data (brown), in data extended with three
  years of weekly follow-up observations (olive) and in three years of
  future data only (dark blue). The shaded regions show the
  frequencies of interest, with color coding as before.}
\label{Fig:FutureData}
\end{figure}

Fig. \ref{Fig:FutureData} shows the upper limits for detection with
the extended data, compared to the upper limits with only the current
data. For low frequencies (e.g., scenarios (A) and (B)), the
sensitivity does not improve with the addition of three-years of
data. A longer monitoring campaign is required for this purpose. For
frequencies relevant to scenario (C), the improvement varies from
marginal to a factor of 2, depending on the precise
frequency. Furthermore, we calculate upper limits based only on the
future data and find that the sensitivity for detection of a secondary
term can reach as low as 20\% of the amplitude of the main component
($\sim$0.03\,mag) at some frequencies.  This suggests that a
multi-year follow-up with frequent sampling, by itself, could
significantly improve the sensitivity to detect high-frequency ($\sim
1$\,yr) peaks; 

\subsection{Implications for SMBH Binary Models}

The lack of a significant secondary peak can place constraints on the
physical parameters of the SMBHB PG1302. In particular, models that
predict two peaks, comparable in height, and separated in frequency by
a factor of few, can be ruled out.  While this is not a generic
prediction of hydrodynamical simulations, for specific mass ratios
within the range of $0.25 < q < 0.45$, two near-equal peaks are
predicted in the periodograms of the mass accretion rate onto the BHs
\citet[][see their Fig. 9]{Farris+2014}.  The specific value of the
disfavoured ratio $q$ depends on disk parameters (viscosity,
temperature) - and also on how one defines the accretion rate. If the
accretion rate is defined based on the total mass accreting into the
central cavity, this $q$ value would be somewhat above $q=0.5$
\citep[see Fig. 6 in][]{2013MNRAS.436.2997D}.

\section{Conclusions}

The presence of multiple periodic components in the variability of SMBHBs is predicted by hydrodynamical simulations of circumbinary disks. The detection of multiple periodicity would provide additional indication that PG1302 is a SMBHB. We analysed the LS periodogram of the optical
light-curve of the SMBHB candidate quasar PG1302 to search for multiple
periodic components, and assessed the statistical significance of
secondary peaks by simulating the variability of PG1302.  Our analysis
returned a peak with FAP=6\%, below our detection threshold.
 By injecting fake
secondary sinusoids into our models, we found that the current data
would only allow the detection of secondary periodic modulations
comparable in amplitude to the main peak. Future observations (for a
few years, at weekly sampling), and/or a more sophisticated analysis,
which can mitigate the effects of aliasing,
could significantly improve the
detection capability and help uncover secondary peaks expected to be
produced by SMBH binaries.

\bibliography{PG1302}

\begin{thebibliography}{32}
\expandafter\ifx\csname natexlab\endcsname\relax\def\natexlab#1{#1}\fi

\bibitem[{{Artymowicz} \& {Lubow}(1994)}]{1994ApJ...421..651A}
{Artymowicz} P., {Lubow} S.~H., 1994, \apj, 421, 651

\bibitem[{{D'Orazio} {et~al}\mbox{.}(2015){D'Orazio}, {Haiman}, {Duffell},
  {Farris}, \& {MacFadyen}}]{Dorazio+2015}
{D'Orazio} D.~J., {Haiman} Z., {Duffell} P., {Farris} B.~D., {MacFadyen} A.~I.,
  2015, \mnras, submitted

\bibitem[{{D'Orazio}, {Haiman} \& {MacFadyen}(2013){D'Orazio}, {Haiman}, \&
  {MacFadyen}}]{2013MNRAS.436.2997D}
{D'Orazio} D.~J., {Haiman} Z., {MacFadyen} A., 2013, \mnras, 436, 2997

\bibitem[{{Drake} {et~al}\mbox{.}(2009){Drake}, {Djorgovski}, {Mahabal},
  {Beshore}, {Larson}, {Graham}, {Williams}, {Christensen}, {Catelan},
  {Boattini}, {Gibbs}, {Hill}, \& {Kowalski}}]{2009ApJ...696..870D}
{Drake} A.~J. {et~al.}, 2009, \apj, 696, 870

\bibitem[{{Eggers}, {Shaffer} \& {Weistrop}(2000){Eggers}, {Shaffer}, \&
  {Weistrop}}]{2000AJ....119..460E}
{Eggers} D., {Shaffer} D.~B., {Weistrop} D., 2000, \aj, 119, 460

\bibitem[{{Farris} {et~al}\mbox{.}(2014){Farris}, {Duffell}, {MacFadyen}, \&
  {Haiman}}]{Farris+2014}
{Farris} B.~D., {Duffell} P., {MacFadyen} A.~I., {Haiman} Z., 2014, \apj, 783,
  134

\bibitem[{{Farris} {et~al}\mbox{.}(2015){Farris}, {Duffell}, {MacFadyen}, \&
  {Haiman}}]{Farris+2015}
{Farris} B.~D., {Duffell} P., {MacFadyen} A.~I., {Haiman} Z., 2015, \mnras,
  447, L80

\bibitem[{{Garcia} {et~al}\mbox{.}(1999){Garcia}, {Sodr{\'e}}, {Jablonski}, \&
  {Terlevich}}]{1999MNRAS.309..803G}
{Garcia} A., {Sodr{\'e}} L., {Jablonski} F.~J., {Terlevich} R.~J., 1999,
  \mnras, 309, 803

\bibitem[{{Graham} {et~al}\mbox{.}(2015){Graham}, {Djorgovski}, {Stern},
  {Glikman}, {Drake}, {Mahabal}, {Donalek}, {Larson}, \&
  {Christensen}}]{2015arXiv150101375G}
{Graham} M.~J. {et~al.}, 2015, Nature, in press, e-print ArXiv:1501.01375

\bibitem[{Graham {et~al}\mbox{.}(2015)Graham, Djorgovski, Stern, Glikman,
  Drake, Mahabal, Donalek, Larson, \& Christensen}]{Graham+2015}
Graham M.~J. {et~al.}, 2015, Nature, 518, 74

\bibitem[{{Grindlay} {et~al}\mbox{.}(2009){Grindlay}
  {et~al.}}]{2009ASPC..410..101G}
{Grindlay} J., {et~al.}, 2009, in ASP Conf. Series, Vol. 410, Preserving
  Astronomy's Photographic Legacy: Current State and the Future of North
  American Astronomical Plates, {Osborn} W., {Robbins} L., eds., p. 101

\bibitem[{{Haehnelt} \& {Kauffmann}(2002)}]{2002MNRAS.336L..61H}
{Haehnelt} M.~G., {Kauffmann} G., 2002, \mnras, 336, L61

\bibitem[{{Haiman}, {Kocsis} \& {Menou}(2009){Haiman}, {Kocsis}, \&
  {Menou}}]{2009ApJ...700.1952H}
{Haiman} Z., {Kocsis} B., {Menou} K., 2009, \apj, 700, 1952

\bibitem[{{Hayasaki}, {Mineshige} \& {Sudou}(2007){Hayasaki}, {Mineshige}, \&
  {Sudou}}]{2007PASJ...59..427H}
{Hayasaki} K., {Mineshige} S., {Sudou} H., 2007, \pasj, 59, 427

\bibitem[{{Henden} {et~al}\mbox{.}(2012){Henden}, {Levine}, {Terrell}, {Smith},
  \& {Welch}}]{2012JAVSO..40..430H}
{Henden} A.~A., {Levine} S.~E., {Terrell} D., {Smith} T.~C., {Welch} D., 2012,
  Journal of the American Association of Variable Star Observers (JAAVSO), 40,
  430

\bibitem[{{Ivezi{\'c}} {et~al}\mbox{.}(2014){Ivezi{\'c}}, {Connolly},
  {Vanderplas}, \& {Gray}}]{astroMLText}
{Ivezi{\'c}} {\v Z}., {Connolly} A., {Vanderplas} J., {Gray} A., 2014,
  Statistics, Data Mining and Machine Learning in Astronomy. Princeton
  University Press

\bibitem[{{Kelly}, {Bechtold} \& {Siemiginowska}(2009){Kelly}, {Bechtold}, \&
  {Siemiginowska}}]{2009ApJ...698..895K}
{Kelly} B.~C., {Bechtold} J., {Siemiginowska} A., 2009, \apj, 698, 895

\bibitem[{{Kormendy} \& {Ho}(2013)}]{kormendy13}
{Kormendy} J., {Ho} L.~C., 2013, \araa, 51, 511

\bibitem[{{Lomb}(1976)}]{1976Ap&SS..39..447L}
{Lomb} N.~R., 1976, \apss, 39, 447

\bibitem[{{MacFadyen} \& {Milosavljevi{\'c}}(2008)}]{2008ApJ...672...83M}
{MacFadyen} A.~I., {Milosavljevi{\'c}} M., 2008, \apj, 672, 83

\bibitem[{{Mahabal} {et~al}\mbox{.}(2011){Mahabal}, {Djorgovski}, {Drake},
  {Donalek}, {Graham}, {Williams}, {Chen}, {Moghaddam}, {Turmon}, {Beshore}, \&
  {Larson}}]{2011BASI...39..387M}
{Mahabal} A.~A. {et~al.}, 2011, Bulletin of the Astronomical Society of India,
  39, 387

\bibitem[{{Noble} {et~al}\mbox{.}(2012){Noble}, {Mundim}, {Nakano}, {Krolik},
  {Campanelli}, {Zlochower}, \& {Yunes}}]{2012ApJ...755...51N}
{Noble} S.~C., {Mundim} B.~C., {Nakano} H., {Krolik} J.~H., {Campanelli} M.,
  {Zlochower} Y., {Yunes} N., 2012, \apj, 755, 51

\bibitem[{{Pojmanski}(1997)}]{1997AcA....47..467P}
{Pojmanski} G., 1997, \actaa, 47, 467

\bibitem[{{Robertson} {et~al}\mbox{.}(2006){Robertson}, {Bullock}, {Cox}, {Di
  Matteo}, {Hernquist}, {Springel}, \& {Yoshida}}]{2006ApJ...645..986R}
{Robertson} B., {Bullock} J.~S., {Cox} T.~J., {Di Matteo} T., {Hernquist} L.,
  {Springel} V., {Yoshida} N., 2006, \apj, 645, 986

\bibitem[{{Roedig} {et~al}\mbox{.}(2012){Roedig}, {Sesana}, {Dotti}, {Cuadra},
  {Amaro-Seoane}, \& {Haardt}}]{2012A&A...545A.127R}
{Roedig} C., {Sesana} A., {Dotti} M., {Cuadra} J., {Amaro-Seoane} P., {Haardt}
  F., 2012, \aap, 545, A127

\bibitem[{{Scargle}(1982)}]{1982ApJ...263..835S}
{Scargle} J.~D., 1982, \apj, 263, 835

\bibitem[{{Sesar} {et~al}\mbox{.}(2011){Sesar}, {Stuart}, {Ivezi{\'c}},
  {Morgan}, {Becker}, \& {Wo{\'z}niak}}]{2011AJ....142..190S}
{Sesar} B., {Stuart} J.~S., {Ivezi{\'c}} {\v Z}., {Morgan} D.~P., {Becker}
  A.~C., {Wo{\'z}niak} P., 2011, \aj, 142, 190

\bibitem[{{Shi} {et~al}\mbox{.}(2012){Shi}, {Krolik}, {Lubow}, \&
  {Hawley}}]{2012ApJ...749..118S}
{Shi} J.-M., {Krolik} J.~H., {Lubow} S.~H., {Hawley} J.~F., 2012, \apj, 749,
  118

\bibitem[{{Springel}, {Di Matteo} \& {Hernquist}(2005){Springel}, {Di Matteo},
  \& {Hernquist}}]{2005MNRAS.361..776S}
{Springel} V., {Di Matteo} T., {Hernquist} L., 2005, \mnras, 361, 776

\bibitem[{{Timmer} \& {Koenig}(1995)}]{1995A&A...300..707T}
{Timmer} J., {Koenig} M., 1995, \aap, 300, 707

\bibitem[{{Vanderplas} {et~al}\mbox{.}(2012){Vanderplas}, {Connolly},
  {Ivezi{\'c}}, \& {Gray}}]{astroML}
{Vanderplas} J., {Connolly} A., {Ivezi{\'c}} {\v Z}., {Gray} A., 2012, in
  Conference on Intelligent Data Understanding (CIDU), pp. 47 --54

\bibitem[{{Zechmeister} \& {K{\"u}rster}(2009)}]{2009A&A...496..577Z}
{Zechmeister} M., {K{\"u}rster} M., 2009, \aap, 496, 577

\end{thebibliography}

\end{document}